\documentclass[conference]{IEEEtran}
\IEEEoverridecommandlockouts
\usepackage{stmaryrd}
\usepackage{amsmath}

\usepackage{flushend}

\usepackage{amsmath}
\usepackage{amsfonts}%
\usepackage{amssymb}
\usepackage{amsthm}
\usepackage{mathtools}
\usepackage{algorithmicx}
\usepackage{algpseudocode}
\usepackage{algorithm}
\usepackage{physics}
\usepackage{relsize}
\usepackage{gensymb}
\usepackage{stmaryrd}
\algdef{SE}[DOWHILE]{Do}{doWhile}{\algorithmicdo}[1]{\algorithmicwhile\ #1}%

\usepackage{cite}  
\usepackage{array}
\usepackage{subcaption}
\usepackage[font=normalsize,labelformat=simple]{subfig}

\usepackage{blindtext}
\usepackage{tikz}
\usepackage{amsthm}

\newtheorem{proposition}{Proposition}

\usepackage[left=0.7in,right=0.7in,top=0.7in,bottom=1in]{geometry}

\begin{document}

\title{Quantum Margulis Codes}

\author{\IEEEauthorblockN{Michele Pacenti
and
Bane Vasi\'c}
\IEEEauthorblockA{Department of Electrical and Computer Engineering, University of Arizona, Tucson, AZ, USA}

mpacenti@arizona.edu, vasic@ece.arizona.edu 

\thanks{This work is supported by the NSF under grants CIF-1855879, CIF-2106189, CCF-2100013, ECCS/CCSS-2027844, ECCS/CCSS-2052751, and in part by the CoQREATE program under grant ERC-1941583 and by Jet Propulsion Laboratory, California Institute of Technology, under a contract with the National Aeronautics and Space Administration and
funded through JPL’s Strategic University Research Partnerships
(SURP) program. Bane Vasi\'{c} has disclosed an outside
interest in his startup company Codelucida to The University
of Arizona. Conflicts of interest resulting from this interest are being managed by The University of Arizona in accordance
with its policies.
}

}

\maketitle

\begin{abstract}
Recently, Lin and Pryadko~\cite{pryadko2024twoblock}  presented the quantum two-block group algebra codes, a generalization of bicycle codes obtained from Cayley graphs of non-Abelian groups. We notice that their construction is naturally suitable to obtain a quantum equivalent of the well-known classical Margulis code. In this paper, we first present an alternative description of the two-block group algebra codes using the left-right Cayley complex; then, we show how to modify the construction of Margulis to get a two-block algebra code. Finally, we construct several quantum Margulis codes and evaluate their performance with numerical simulations.
\end{abstract}


\IEEEpeerreviewmaketitle

\section{Introduction}
%
%
%
%
In the last two years, asymptotically good quantum LDPC codes have finally been shown to exist, and explicit constructions were given in  \cite{panteleev_asymptotically_2022,leverrier_quantum_2022,dinur_good_2023}. However, there are still questions on the applicability of this class of codes in realistic scenarios. Moreover, it is not clear how these codes could perform in the finite length regime. Recently, a class of codes called \textit{bivariate codes} \cite{bravyi2024high} has been constructed and utilized in an experimental setup of fault tolerance. Bivariate codes belong to the wider class of \textit{generalized bicycle (GB)} codes \cite{pryadko2022gb}, a class of quantum LDPC codes obtained from a pair of two circulant blocks. Although GB codes have not the same desirable asymptotic properties of the codes presented in \cite{panteleev_asymptotically_2022,leverrier_quantum_2022,dinur_good_2023}, the experiment of \cite{bravyi2024high} motivates their study, as they are particularly suitable for hardware implementation, and have good minimum distance and code rate for  finite blocklengths.

Recently, GB codes were further generalized into the so-called \textit{two-block group algebra} (2BGA) codes \cite{pryadko2024twoblock}. The parity check matrices of this class of codes are obtained from the Cayley graphs of some group, generated by a given set of generators, and GB codes can be seen as special cases of 2BGA codes. The authors in \cite{pryadko2024twoblock} have derived bounds on the minimum distance, and have enumerated all the 2BGA codes with optimal parameters (up to permutations) for blocklengths less than 200, showing an increase of the minimum distance with the square root of the blocklenght. The considered groups were products and semi-products of cyclic groups and the dihedral group.

We notice that the 2BGA construction is naturally suitable to extend the well-known classical LDPC construction from Margulis \cite{margulis1982explicit}, as well as other similar constructions such as the Margulis-Ramanujan construction \cite{rosenthal_constructions_2001, MACKAY200397}, to obtain quantum LDPC codes. The classical codes obtained from these constructions have rate $R=1/2$, and their parity check matrix is composed by two square blocks corresponding to the incidence matrix of Cayley graphs of the special linear group $SL(2,\mathbb{Z})$ in the case of Margulis codes, and the projective general linear $PGL(2,\mathbb{Z})$ and projective special linear $PSL(2,\mathbb{Z})$ groups in the case of Margulis-Ramanujan codes.

In this paper, we first present an alternative description of 2BGA codes using the left-right Cayley complex. Then, we modify the construction of Margulis to obtain a class of 2BGA codes with blocklength larger than 200, and we show that the same result on the girth of the code of \cite{margulis1982explicit} is applicable to our construction. Finally, we construct several examples of quantum Margulis codes, with different blocklengths, girth and variable/check node degree, and evaluate their performance via numerical simulation under depolarizing noise.

This paper is organized as follows: in Section \ref{sec:premilinaries}, we present the preliminaries and introduce the notation, in Section \ref{sec:description} we describe 2BGA codes over the left-right Cayley complex, in Section \ref{sec:construction} we recall the construction of Margulis and extend it to get quantum Margulis codes and finally, in Section \ref{sec:results} we construct several quantum Margulis codes and show Monte Carlo simulation results for logical error rate.

\section{Preliminaries}
\label{sec:premilinaries}
Let $\mathbb{F}_2^{n}$ be the field of the binary vectors of length $n$; the \textit{Hamming weight} (or simply weight) of an element in $\mathbb{F}_2^{n}$ is the number of its non-zero entries. An $[n,k,d]$ linear code $C \subset \mathbb{F}_2^{n}$ is a linear subspace of $\mathbb{F}_2^{n}$ generated by $k$ elements, such that each element in $C$ has Hamming weight at least $d$. A code $C$ can be represented by an $(n-k) \times n$ parity check matrix $\mathbf{H}$ such that $C = \ker \mathbf{H}$. If $\mathbf{H}$ is \textit{sparse}, i.e., its row and column weights are less than $\log n$, the code $C$ is a \textit{low-density parity check} code. A graph $\Gamma = (V,E)$ is a collection of vertices $V$ and edges $E$, such that each edge connects two distinct vertices $v_i,v_j$ and can be represented by the pair $(v_i,v_j)$. To a parity check matrix is associated a bipartite graph called \textit{Tanner graph} $\mathcal{T} = (V \cup C, E)$ \cite{tanner_recursive_1981}, where the nodes in $V$ are called \textit{variable nodes}, the nodes in $C$ are called \textit{check nodes}, and there is an edge between $v_j \in V$ and $c_i \in C$ if $h_{ij}=1$, where $h_{ij}$ is the element in the $i$-th row and $j$-th column of $\mathbf{H}$. The \textit{degree} of a node is the number of incident edges to that node. If all the variable (check) nodes have the same degree we say the code has \textit{regular} variable (check) degree, and we denote it with $d_v$ ($d_c)$. A \textit{cycle} is a closed path in the Tanner graph, and we denote its length by the number of variable and check nodes in the cycle. The \textit{girth} $g$ of a Tanner graph is the length of its shortest cycle. 

Given a group $G$ and a set of generators $S$, it is possible to construct the Cayley graph $\mathrm{Cay}(G,S)$ of $G$ with respect of $S$, such that $\mathrm{Cay}(G,S) = (G,E_S)$ is a graph where there is a vertex for every element $g \in G$, and there is an edge for every pair $(g,gs)$, with $s\in S$, if $S$ acts on the right. Alternatively, if $S$ acts on the left, the edges have the form $(g,sg)$.

Let $(\mathbb{C}^2)^{\otimes n}$ be the $n$-dimensional Hilbert space, and $P_n$ be the $n$-qubit Pauli group; a \textit{stabilizer} group is an Abelian subgroup $S \subset P_n$, and an $\llbracket n,k,d \rrbracket$ stabilizer code  is a $2^k$-dimensional subspace $\mathcal{C}$ of $(\mathbb{C}^2)^{\otimes n}$ that satisfies the condition $s_i\ket{\Psi} = \ket{\Psi},\ \forall\ s_i\in S, \ket{\Psi}\in \mathcal{C}$. An $\llbracket n, k_X-k_Z, d \rrbracket $ Calderbank-Shor-Steane (CSS) code $\mathcal{C}$ is a stabilizer code constructed using two classical  $[n,k_X,d_X]$ and $[n,k_Z,d_Z]$ codes $C_X = \ker \mathbf{H}_X$ and $C_Z = \ker \mathbf{H}_Z$, respectively, where $d\geq~\mathrm{min}\{d_X,d_Z\}$ and $C_Z \subset C_X$ \cite{calderbank_good_1996}. Note that $k_X$, $k_Z$  and $d_X$, $d_Z$ correspond to the dimensions and minimum distances of $C_Z$ and $C_X$, respectively. 
 
A chain complex 
$$
\cdots \xrightarrow{\partial_{i+1}} \mathcal{C}_i \xrightarrow{\partial_{i}} \mathcal{C}_{i-1} \xrightarrow{\partial_{i-1}} \cdots
$$
is a sequence of abelian groups and morphisms called \textit{boundary maps} such that $\partial_i \circ \partial_{i+1} = 0$ for all $i \in \mathbb{Z}$ \cite{panteleev_asymptotically_2022}. This property implies that $\mathrm{im}\partial_{i+1} \subseteq \mathrm{ker}\partial_i$, thus one can consider the quotient group $H_i(\mathcal{C}) = \mathrm{ker}\partial_i / \mathrm{im}\partial_{i+1}$, called the \textit{i-th homology group} of $\mathcal{C}$. Alternatively, it is possible to define a co-chain complex
$$
\cdots \xleftarrow{\partial^{i+1}} \mathcal{C}^{i+1} \xleftarrow{\partial^{i}} \mathcal{C}^{i} \xleftarrow{\partial^{i-1}} \cdots
$$
to be the dual of a chain complex. Here the morphisms $\partial^i$ are called \textit{co-boundary maps} and $\partial^{i+1} \circ \partial^{i} = 0$, which is equivalent to $\mathrm{im}\partial^{i} \subseteq \mathrm{ker}\partial^{i+1}$, allows us to consider the quotient group $H^i(\mathcal{C}) =  \mathrm{ker}\partial^{i+1} / \mathrm{im}\partial^{i} $ called the \textit{i-th cohomology group}.

A classical linear code can be interpreted as a 2-term chain complex $\mathcal{C} : \mathbb{F}_2^n \xrightarrow[]{\mathbf{H}} \mathbb{F}_2^{n-k}$ such that its first homology group $H_1(\mathcal{C})$ corresponds to $\ker \mathbf{H}$. 

A quantum CSS code can be represented by the 3-term chain complex $\mathcal{C}: \mathbb{F}_2^{m_x}\xrightarrow{\partial_2} \mathbb{F}_2^{n} \xrightarrow{\partial_1}  \mathbb{F}_2^{m_z} $, with $\partial_2 \in \mathbb{F}_2^{m_x\times n}$ and  $\partial_1 \in \mathbb{F}_2^{n\times m_z}$ being the first and the second boundary maps, respectively; the space $\mathbb{F}_2^{m_x}$ (resp. the space $\mathbb{F}_2^{m_z}$) corresponds to the space of the $Z$-checks (resp. the $X$-checks), while the space $\mathbb{F}_2^{n} $correspond to the space of the $n$ qubits. Alternatively, we can represent the quantum code by its dual chain $\mathcal{C}^{*}: \mathbb{F}_2^{m_x}\xleftarrow{\partial^1} \mathbb{F}_2^{n} \xleftarrow{\partial^2}  \mathbb{F}_2^{m_z} $. 
We can identify the code $C_Z$ with the subcomplex $\mathbb{F}_2^{n} \xrightarrow{\partial_1}  \mathbb{F}_2^{m_z}$, and its dual $C_Z^{\perp}$ with the subcomplex $\mathbb{F}_2^{n} \xleftarrow{\partial^2}  \mathbb{F}_2^{m_z} $; similarly, we can identify $C_X$ with the subcomplex $\mathbb{F}_2^{m_x}\xleftarrow{\partial^1} \mathbb{F}_2^{n}$, and its dual $C_X^{\perp}$ with the subcomplex $\mathbb{F}_2^{m_x}\xrightarrow{\partial_2} \mathbb{F}_2^{n} $. It follows that the length of the quantum code is equal to $n$, and its dimension $k$ is the dimension of the first homology group $H_1(\mathcal{C})$ (or of its first cohomolgy group $H^1(\mathcal{C^*})$).
We naturally identify $\mathbf{H}_X \triangleq \partial^1$ and  $\mathbf{H}_Z \triangleq \partial_1$\footnote{Note that $\partial^1 = \partial_2^T$ and $\partial^2 = \partial_1^T$.}.

\section{Alternative description of the two-block group algebra codes}
\label{sec:description}
In this section we present an alternative description of the two-block group algebra codes using the left-right Cayley complex proposed in \cite{dinur2022ltc}. This representation highlights the strong relation between this class of codes and other classes of quantum LDPC codes: it is known, for instance, that two-block group algebra codes are the smallest lifted product codes \cite{pryadko2024twoblock}. Although a description of lifted product codes on the left-right Cayley complex is exploited in \cite{panteleev_asymptotically_2022}, there is no such a description of bicycle codes and two-block group algebra codes. Moreover, quantum Tanner codes \cite{leverrier_quantum_2022} are also defined on a left-right Cayley complex, with the only difference on how qubits and checks are assigned over the graph. Thus, our goal is to fill this gap by showing that it is possible to define two-block group algebra codes on left-right Cayley complexes as well, as this may enable new methodologies for the study of these codes such as expander arguments, as done in \cite{panteleev_asymptotically_2022,leverrier_quantum_2022}.

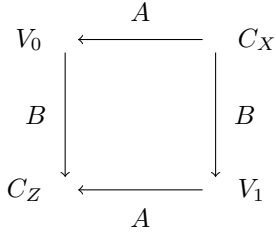
\begin{figure}
\centering
    \begin{tikzpicture}
\tikzstyle{var} = [circle, draw=white]

\node[var, label=left:$C_Z$] (cz) at (0,0){};
\node[var, label=right:$V_1$] (v0) at (2,0){};
\node[var, label=left:$V_0$] (v1) at (0,2){};
\node[var, label=right:$C_X$] (cx) at (2,2){};

\draw [->] (cx)--(v0) node[midway, label=right:$B$]{};
\draw [->] (v0)--(cz) node[midway, label=below:$A$]{};
\draw [->] (cx)--(v1) node[midway, label=above:$A$]{};
\draw [->] (v1)--(cz) node[midway, label=left:$B$]{};
\end{tikzpicture}
    \caption{Left-right Cayley complex with code assignments.}
    \label{fig:lrcayley2}
\end{figure}

We begin by summarizing the construction of the traditional bicycle codes. Let $\mathbf{A},\mathbf{B} \in \mathbb{F}_2^{\ell \times \ell}$ be two circulant matrices. The CSS code is defined by the two parity check matrices
\begin{equation}
\mathbf{H}_X = [\mathbf{A},\mathbf{B}],\ \mathbf{H}_Z = [\mathbf{B}^T,\mathbf{A}^T],
\end{equation}
and has length $n=2\ell$. Because of the fact that circulant matrices always commute, we have
\begin{equation}
    \mathbf{H}_X\mathbf{H}_Z^T = [\mathbf{A},\mathbf{B}]\cdot \begin{bmatrix}
        \mathbf{B} \\ \mathbf{A}
    \end{bmatrix} = \mathbf{AB} + \mathbf{BA} = \mathbf{0}.
\end{equation}
Let now $G$ be a group and $A,B \subset G$ two sets of generators, $A$ acting on the left and $B$ acting on the right. Let us recall the construction of the \textit{left-right Cayley complex} associated to $G$ \cite{dinur2022ltc}, in its quadripartite version. We take 4 copies of $G$, and we call them respectively $V_0,V_1,C_X,C_Z$. Next, we construct $\Gamma_{A_1} = (C_X \cup V_0, E_A)$, $\Gamma_{A_2} = (C_Z \cup V_1, E_A)$, $\Gamma_{B_1} = (C_X \cup V_1, E_B)$ and $\Gamma_{B_2} = (C_Z \cup V_0, E_B)$, such that $\Gamma_{A_1}$ is isomorphic to $\Gamma_{A_2}$, and $\Gamma_{B_1}$ is isomorphic to $\Gamma_{B_2}$, with $\Gamma_{A_1}$ and $\Gamma_{B_1}$ being double covers of the Cayley graph of $G$ with respect to $A$ and $B$. The left-right Cayley complex is obtained as  $\boldsymbol{\Gamma} = \Gamma_{A_1} \cup \Gamma_{A_2} \cup \Gamma_{B_1} \cup \Gamma_{B_2}$. Note that, in this paper, we do not assume $A^{-1} = A$ and $B^{-1}=B$, thus the constructed Cayley graphs may be digraphs. The complex can be visualized as illustrated in Fig.\ref{fig:lrcayley2}.

We can now associate a chain complex to the left-right Cayley complex we have constructed. Let $X$ be a 3-term chain complex over the group algebra $\mathbb{F}_2G$ such that
\begin{equation}
        X : \mathbb{F}_2^{C_X} \xrightarrow{\partial_2} \mathbb{F}_2^{V_0} \oplus \mathbb{F}_2^{V_1} \xrightarrow{\partial_1}  \mathbb{F}_2^{C_Z},
\end{equation}
with
\begin{equation}
    \begin{aligned}
        \partial_2 \triangleq &\ (ac_X,c_Xb), \forall c_X\in {F}_2^{C_X} \\
        \partial_1 \triangleq &\ v_0b+ av_1 , \forall v_0\in \mathbb{F}_2^{V_0}, v_1\in \mathbb{F}_2^{V_1},\\
    \end{aligned}
\end{equation}
where with $\mathbb{F}_2^G$ we denote the space of all the formal linear combinations of elements of $G$ with binary coefficients.
In other words, $\partial_2$ is a map between $C_X$ and the direct sum of $V_0$ and $V_1$, such that each element in its image is a pair of elements of $G$, and $\partial_1$ is a map between $V_0 \oplus V_1$ and $C_Z$ such that each element in its image is the sum over $\mathbb{F}_2G$ of the two elements in its input. It is straightforward to see that $X$ is a well-defined chain complex, \textit{i.e.}, $ \partial_1 \circ \partial_2=0$. Indeed, for any $c_X \in C_X$ we first compute $\partial_2(c_X) = (ac_X,c_Xb)$, then we compute $\partial_1(\partial_2(c_X)) = \partial_1(ac_X,c_Xb) = ac_Xb + ac_Xb = 0$.
We notice that $\partial_2$ and $\partial_1$ have a natural binary representation, which we denote with $\boldsymbol{\partial}_2$ and $\boldsymbol{\partial}_1$, respectively, such that  $\boldsymbol{\partial}_2=[\mathbf{A},\mathbf{B}]$, where $\mathbf{A},\mathbf{B} \in \mathbb{F}_2^{|G|\times |G|}$ are the biadjacency matrices of the double covers of the Cayley graphs of $G$ in respect of $A$ and $B$, respectively, and $\boldsymbol{\partial}_1 = [\mathbf{B},\mathbf{A}]^T$. Similarly we can consider the cochain complex
\begin{equation}
            X^* : \mathbb{F}_2^{C_X} \xleftarrow{\partial^1} \mathbb{F}_2^{V_0} \oplus \mathbb{F}_2^{V_1} \xleftarrow{\partial^2}  \mathbb{F}_2^{C_Z}, 
\end{equation}
such that $\boldsymbol{\partial}^1 = [\mathbf{A}^T,\mathbf{B}^T]^T$ and $\boldsymbol{\partial}^2 = [\mathbf{B}^T,\mathbf{A}^T]$. Note that $\mathbf{A}^T,\mathbf{B}^T$ correspond to the incidence matrices of the double covers of the Cayley graphs of $G$ in respect of $A^{-1}$ and $B^{-1}$, respectively; indeed, we can represent the cochain complex $X^*$ by reversing the arrows in Fig. \ref{fig:lrcayley2}, and substituting $a$ and $b$ with $a^{-1}$ and $b^{-1}$. Because the corresponding Cayley graphs are directed, the biadjacency matrices may not be symmetric.

We are now able to associate a quantum CSS code to the complex, such that $\mathbf{H}_X = \boldsymbol{\partial}_2$, $\mathbf{H}_Z = \boldsymbol{\partial}^2$.
Having defined $A$ and $B$ acting respectively on the left and on the right, we now have that their action commute, thus
\begin{equation*}
    \mathbf{H}_X\mathbf{H}_Z^T = [\mathbf{A},\mathbf{B}] \cdot [\mathbf{B},\mathbf{A}]^T = \mathbf{A}\mathbf{B} + \mathbf{B}\mathbf{A} = \mathbf{0}.
\end{equation*}


\section{Quantum Margulis codes}
\label{sec:construction}
In this section, we recall the construction of classical LDPC codes from Margulis, and show how it can be adapted to construct quantum Margulis codes. Moreover, we show that the same result on the girth of the code can be applied to the quantum case.
\subsection{Classical Margulis construction}
Margulis codes \cite{margulis1982explicit} are a well-known class of classical LDPC codes constructed from Cayley graphs of certain groups. 
Let $G$ be $SL(2,p)$ the \textit{Special Linear Group} whose elements consist of $2\times 2$ matrices of determinant 1 over $\mathbb{Z}_p$, being $p$ prime. Let $S$ be a set of generators of $SL(2,\mathbb{Z}_p)$ chosen according to the construction we report below, and let $S^{-1}$ be the inverse of $S$. Let $\Gamma = (G\times \mathbb{F}_2 \cup G,E)$ be a bipartite graph, with the set of left vertices being two distinct copies of $G$, and the set of right vertices to be $G$. The subgraph $\Gamma_0 = (G\times 0 \cup G,E_S)$ is the Cayley graph of $G$ with respect of the generators $S$, and the subgraph $\Gamma_1 = (G\times 1 \cup G,E_{S^{-1}})$ is the Cayley graph of $G$ with respect of the generators $S^{-1}$. The LDPC code is constructed by assigning the left vertices to be bits and the right vertices to be checks; it has blocklength $n = 2(p^2-1)p$, rate $R=1/2$, variable degree $d_v=|S|$ and check degree $d_c=2|S|$.
A similar construction is based on Ramanujan graphs by Lubotzky \textit{et al.} \cite{rosenthal2000,MACKAY200397}, however we will only consider Margulis construction for this paper.

Margulis shows that if the set of generators are chosen such that there is no multiplicative relation between them, the girth of the graph grows as $\log p$; moreover, he gives an explicit construction of the generating set for any degree satisfying this property. Let $\eta$ be a sufficiently large integer, and let us select $r+1$ distinct pairs $(m_i,q_i)$, with $1 \leq i \leq r+1$, such that $\gcd (m_i,q_i) = 1$ and $0 \leq m_i \leq \eta/2$, $0 \leq q_i \leq \eta/2$. For each pair, there exist a matrix
\begin{equation}
    C_i = \begin{pmatrix}
        m_i & a_i \\
        q_i & b_i
    \end{pmatrix} \in SL(2,\mathbb{Z})
\end{equation}
such that $|a_i|,|b_i| < \eta/2$. Each generator $g_i$ is then given by 
\begin{equation}
    g_i = C_i \begin{pmatrix}
        1 & \eta \\
        0 & 1
    \end{pmatrix}C_i^{-1}, \forall\ i=1,..,r+1.
\end{equation}
The generating set is defined as $S = \{g_1,...,g_r\}$, with $S^{-1}=\{g_1^{-1},...,g_r^{-1}\}$. Because each $g_i \in SL(2,\eta\mathbb{Z})$, Margulis shows that there exist no nontrivial multiplicative relation between the generators, and that the girth of the code grows as $\mathcal{O}(\log n / \log r)$. Generally, it is sufficient to choose $\eta< \sqrt{7r}$.

\subsection{Quantum Margulis construction}
The construction from Margulis can be extended to fit the design proposed in Section \ref{sec:construction}. Let $G=SL(2,\mathbb{Z}_p)$, with $p$ prime. We use Margulis' method to obtain a set $S$ of $r$ generators and its inverse $S^{-1}$; for simplicity, let us assume $r$ even, so that we can divide $S$ in two subgroups of equal size $r/2$, such that $S = \{A,B\}$; similarly we can split $S^{-1} = \{A^{-1},B^{-1}\}$. Let $A$ act on the left and $B$ act on the right. We construct the left-right Cayley complex as in Section \ref{sec:construction}, and associate the correspondent quantum code. The structure of the parity check matrices $\mathbf{H}_X,\mathbf{H}_Z$ is very similar to the one of the classical Margulis code. Qubits are associated with two distinct copies of $G$ (namely, $V_0$ and $V_1$), and $X,Z$ checks are associated with a copy of $G$ (namely, $C_X$ and $C_Z$, respectively). The graph $\Gamma_X = ((V_0 \cup V_1) \cup C_X,E_X)$ can be subdivided into two subgraphs: $\Gamma_{X0} = (V_0 \cup C_X,E_A)$, which is the Cayley graph of $G$ with respect of $A$ acting on the left, and $\Gamma_{X1} = (V_1 \cup C_X,E_B)$, which is the Cayley graph of $G$ with respect of $B$ acting on the right. Similarly, the graph $\Gamma_Z = ((V_0 \cup V_1) \cup C_Z,E_Z)$ can be subdivided into two subgraphs: $\Gamma_{Z0} = (V_0 \cup C_Z,E_{B^{-1}})$, which is the Cayley graph of $G$ with respect of $B^{-1}$ acting on the right, and $\Gamma_{Z1} = (V_1 \cup C_Z,E_{A^{-1}})$, which is the Cayley graph of $G$ with respect of $A^{-1}$ acting on the left. The parity check matrices $\mathbf{H}_X,\mathbf{H}_Z$ are the incidence matrices of $\Gamma_X,\Gamma_Z$, respectively. 
The quantum code has length \mbox{$n=2(p^2-1)p$}, regular variable degree $d_v=r$, and regular check degree $d_c=2r$. Because both $\mathbf{H}_X$ and $\mathbf{H}_Z$ have $|G|$ rows and $2|G|$ columns, the rate of the quantum code $R_q \rightarrow 0$ for $n\rightarrow \infty$; however, because they always have redundant rows, we are able to construct finite length codes with non-trivial dimension\footnote{Note that this is true for $SL(2,\mathbb{Z}_q)$, but it's not true in general.}. We reserve to improve the rate of these codes in future work.

Because the structure of the code is essentially unchanged, we show that Margulis' argument for the girth can also be applied to our construction.
\begin{proposition}
\label{theo:girth}
    The girth of the two parity check matrices of the quantum Margulis code increases as $\mathcal{O}(\log n/\log 2d_c)$.
    \begin{proof}
        By design we have that $s\in SL(2,\eta\mathbb{Z})$, for each $s\in S$. Because of this property, Margulis showed that the Cayley graph $\mathrm{Cay}(G,S\cup S^{-1})$ has girth increasing as $\mathcal{O}(\log n/\log d_c)$, with $d_c = r/2$. In our construction, in order to have matrix $\mathbf{H}_X$ to have the same degrees as the classical construction, we need two sets of generators $A,B$ both of cardinality $r$, which means that the total generating set is actually of size $2r$, and $d_c=r$. Let us now consider $\mathrm{Cay}(G,A\cup B \cup A^{-1} \cup B^{-1})$; notice that the fact that $A$ acts on the left is irrelevant, as the Cayley graph $(G,A)$ with $A$ acting on the left is isomorphic to the Cayley graph $(G,A)$ with $A$ acting on the right via the map $g \mapsto g^{-1}$ \cite{dinur2022ltc}. Thus we can apply the proof of Margulis to $\Gamma$ and get that its girth increases as $\mathcal{O}(\log n/\log 2d_c)$.


        
    \end{proof}
\end{proposition}

\section{Numerical results}

\begin{table}
    \centering
    \begin{tabular}{|c||c|c|c|c|c|}
    \hline 
     Code &  $n$  & $k$ & $d_v$ & $d_c$ & $g$ \\
       \hline 
       P5G8D5  & 240 & 8 & \{2,3\} & 5 & 8 \\
       \hline 
       P7G8D5 & 672 & 4 & \{2,3\} & 5 & 8 \\ 
       \hline 
       P11G8D5 & 2640 & 4 & \{2,3\} & 5 & 8 \\
       \hline 
       P7G6D6 & 672 & 10 & 3 & 6 & 6 \\ 
       \hline 
       P7G6D7 & 672 & 6 & \{3,4\} & 7 & 6 \\ 
       \hline 
       P7G6D8 & 672 & 4 & 4 & 8 & 6\\
       \hline 
    \end{tabular}
    \caption{Parameters of the quantum Margulis codes.}
    \label{tab:codes}
\end{table}

\label{sec:results}
In this section, we utilize the construction described in Section \ref{sec:construction} to obtain several quantum Margulis codes. For each code, we perform a Monte Carlo simulation of decoding with BP-OSD with exhaustive search of order 10 \cite{roffe_decoding_2020}, under depolarizing noise. The decoder runs for a maximum number of iterations equal to the blocklenght of the code. For each probability of error, $10^4$ error patterns are simulated; if, after $10^4$ errors, less than 100 logical errors are detected, the simulation continues with the same probability of error until 100 logical errors are detected. We generate several codes with different blocklength, dimension, variable/check degree, and girth, as depicted in Table \ref{tab:codes}. The nomenclature we use for each code is in the form "P\#G\#D\#", where the number following "P" is the prime $p$ of the group $SL(2,p)$ used to construct the code, the number following "G" is the girth, and the number following "D" is the check degree.

\begin{figure}
    \centering
    \includegraphics[scale=.65]{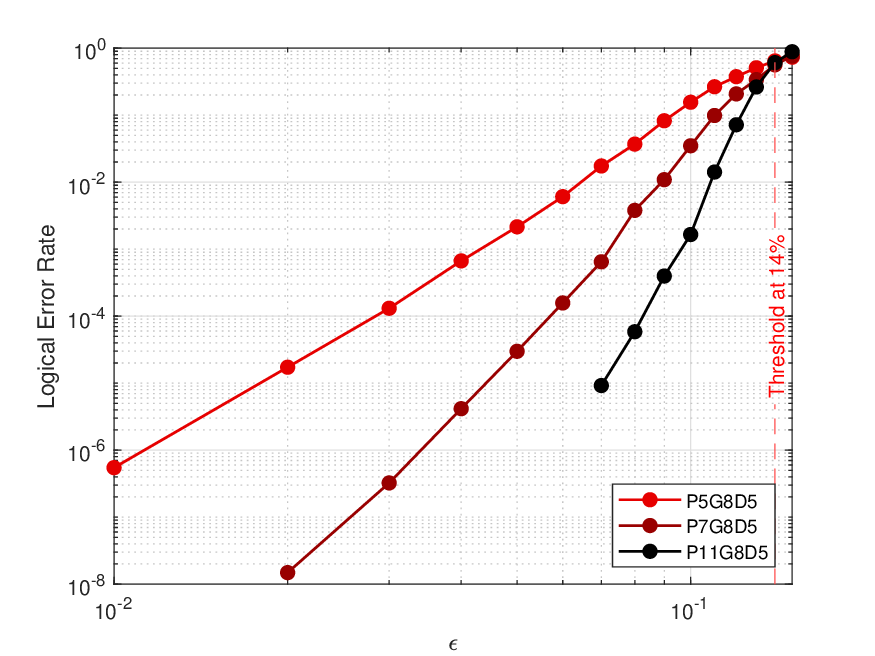}
    \caption{Decoding simulation of three quantum Margulis codes with check degree 5 and increasing blocklength.}
    \label{fig:best_codes}
\end{figure}

\begin{figure}
    \centering
    \includegraphics[scale=.65]{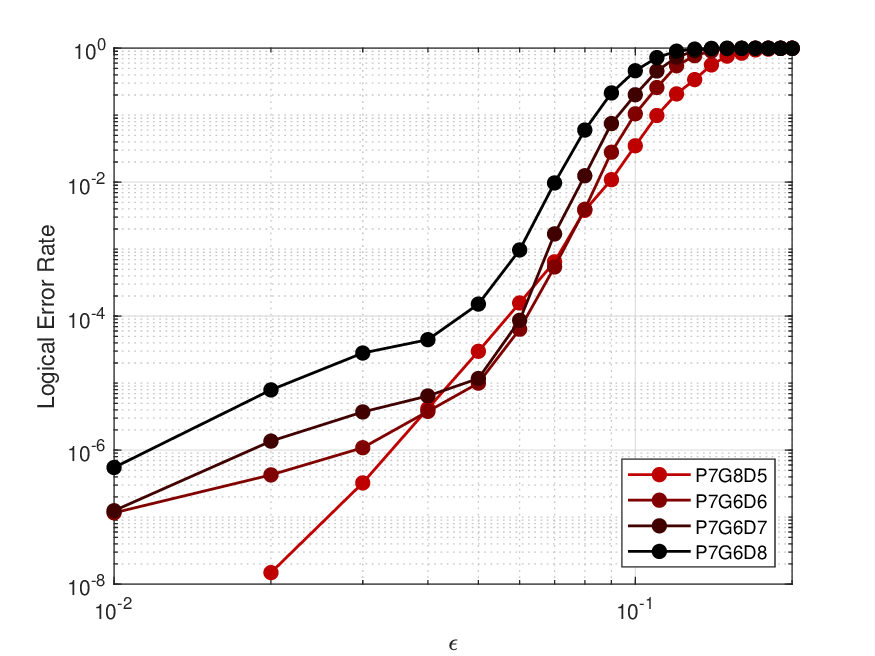}
    \caption{Comparison of quantum Margulis codes of length 672 and increasing check degree.}
    \label{fig:degrees}
\end{figure}

In Fig. \ref{fig:best_codes} we illustrate the performance of several codes constructed with $p=5,7,11$; all the codes have girth $g=8$, regular check degree $d_c=5$, and two sets of $n/2$ variable nodes, one set with $d_v=2$ and the other with $d_v=3$. Although we also have designed codes with $p=13$, its blocklength of $n=4368$ makes it infeasible to decode with BPOSD. We reserve for future study the analysis of longer codes. We were able to construct several codes with girth $g=8$ and check degree 5 that perform well under BPOSD decoding, giving rise to a decoding threshold around $14\%$ (we stress that the terminology "threshold" here is slightly abused, as each code has a different code rate). We reserve for future study the performance of these codes under different decoders, ideally with lower decoding complexity. We also stress that, for this paper, we focused on codes with $d_v = \{2,3\}$ and $d_c=5$, because we find that they generally have higher girth than, for instance, $d_v=3$ and $d_c=6$, and better decoding threshold. Nevertheless, is is likely that codes in the $(d_v=3,d_c=6)$ ensemble will have better distance properties. In Fig. \ref{fig:degrees}, for instance, we fix the code length, and construct several codes with increasing check degree. Specifically, we have codes with check degree from 5 to 8. As mentioned earlier, the code with check degree 5, namely P7G8D5, shows the best performance both in terms of threshold and error floor; however, the code P7G6D6 seems to have a better slope, which may be due to better minimum distance, although its threshold is lower, and its girth $g=6$ is also lower, thus giving rise to an error floor. We also simulate P7G6D7, which has $(d_v=\{3,4\},d_c=7)$, and P7G6D8, which has $(d_v=4,d_c=8)$, and we observe that the threshold decreases while increasing the variable and check degrees. Unfortunately, we weren't able to find codes in these ensembles with $g=8$, and we reserve this task for future work.

\section{Conclusion}
\label{sec:conclusion}
We constructed 2BGA codes of blocklength $n>200$ by modifying the construction of Margulis for classical LDPC codes. A similar "generalization" can be applied to the class of Margulis-Ramanujan classical LDPC codes \cite{rosenthal_constructions_2001}; however, we were able to design codes with high girth only for blocklengths of $\approx 5000$, which is too high to be decoded efficiently with the BPOSD decoder. Thus, we reserve the study of these codes for future work.
We obtained new quantum LDPC with moderate blocklengths and analyzed their performance under BPOSD decoding. We also extended the result of Margulis on the girth of the code. A particular subset of quantum Margulis codes with check degree of 5 shows excellent performance both in the waterfall and error floor region; however, the code rate goes to 0 with $n\rightarrow \infty$, as happens in general for 2BGA codes. It would be interesting to find ways to improve the code rate while minimally impacting the minimum distance of the code. Another interesting application of quantum Margulis codes (and quantum Margulis-Ramanujan codes) is in the context of quantum expanders \cite{panteleev_asymptotically_2022,leverrier_quantum_2022,dinur_good_2023,leverrier2015expander}, which have been used to design asymptotically good quantum LDPC codes.



\ifCLASSOPTIONcaptionsoff
  \newpage
\fi



\bibliographystyle{IEEEtran}


%

\end{document}